\documentclass{JAC2001}                                                  
\usepackage{graphicx}                                                          
\newcommand{\ud}{\mathrm{d}}                                                                                                                                                                                                               
\begin{document}                                                               
                                                                               
\title{QUASICLASSICAL CALCULATIONS OF  WIGNER FUNCTIONS IN      
NONLINEAR BEAM DYNAMICS}                                                         
\author{Antonina N.  Fedorova, Michael G. Zeitlin\\
IPME, RAS, V.O. Bolshoj pr., 61, 199178, St.~Petersburg, Russia                
\thanks{e-mail: zeitlin@math.ipme.ru}\thanks{ http://www.ipme.ru/zeitlin.html; 
http://www.ipme.nw.ru/zeitlin.html}
}

\maketitle                          

\begin{abstract} 
We present the application of variational-wavelet analysis to                  
numerical/analytical calculations of Wigner functions in (nonlinear)           
quasiclassical beam dynamics problems. (Naive) deformation                     
quantization and multiresolution representations are the key points. We        
construct the representation via multiscale expansions in generalized          
coherent states or high-localized nonlinear eigenmodes in the base of           
compactly supported wavelets and wavelet packets.

\end{abstract}

\section{INTRODUCTION}

In this paper we consider the  applications of a new 
nu\-me\-ri\-cal\--\-ana\-ly\-ti\-cal technique based on local nonlinear harmonic analysis
(wavelet analysis, generalized coherent states analysis) to  quantum/quasiclassical
(nonlinear) beam/accelerator physics calculations. The reason for this treatment is that
recently a number of problems appeared in which one needs take into account quantum properties
of particles/beams. Our starting point is the general point of view of deformation quantization approach at least on
naive Moyal/Weyl/Wigner level [1], [2].

The main point is that the algebras of quantum observables are the deformations
of commutative algebras of classical observables (functions) [1]. So, if we have the Poisson manifold $M$ (symplectic 
manifolds, Lie coalgebras, etc) as a model for classical dynamics then for quantum calculations we need to find
an associative (but non-commutative) star product $*$ on the space of formal power series in $\hbar$ with
coefficients in the space of smooth functions on $M$ such that
\begin{eqnarray}
f * g =fg+\hbar\{f,g\}+\sum_{n\ge 2}\hbar^n B_n(f,g), 
\end{eqnarray}
where
$\{f,g\}$
is the Poisson brackets, $B_n$ are bidifferential operators.
Kontsevich gave the solution to this deformation problem in terms of formal
power series via sum over graphs and proved that for every Poisson manifold M there is a canonically
defined gauge equivalence class of star-products on M. Also there are nonperturbative corrections to power 
series representation for $*$  [1]. In naive calculations we may use simple formal rule:
\begin{eqnarray}
* &\equiv&\exp \Big(\frac{i\hbar}{2}(\overleftarrow\partial_x\overrightarrow\partial_p-
   \overleftarrow\partial_p\overrightarrow\partial_x)\Big)
\end{eqnarray}
In this paper we consider calculations of Wigner functions (WF) as the solution
of Wigner equations [2] (part 2):

\begin{eqnarray}
i\hbar\frac{\partial}{\partial t}W(x,p,t)=H * W(x,p,t)-W(x,p,t) * H
\end{eqnarray}
and especially stationary Wigner equations.
Our approach is based on extension of our variational-wavelet approach [3]-[14].
Wavelet analysis is some set of mathematical methods, which gives us the possibility to
work with well-localized bases in functional spaces and gives maximum sparse
forms for the general type of operators (differential, integral, pseudodifferential) in such bases.
These bases are natural generalization of standard coherent, squeezed, thermal squeezed states [2],
which correspond to quadratical systems (pure linear dynamics) with Gaussian Wigner functions.
So, we try to calculate quantum corrections to classical dynamics described by polynomial nonlinear 
Hamiltonians such as orbital motion in storage rings, orbital dynamics in general multipolar
fields  etc. from papers [3]-[14].
The common point for classical/quantum calculations is that
any solution, 
which comes from full multiresolution expansion in all space/time (or phase space)
scales, is represented via expansion into a slow part
and fast oscillating parts (part 3).
So, we may move
from the coarse scales of resolution to the 
finest one to obtain  more detailed information about our dynamical classical/quantum process.
In this way we give contribution to our full solution
from each scale of resolution. 
The same is correct for contributions to power spectral density
(energy spectrum): we can take into account contributions from each
level/scale of resolution.
Because affine
group of translations and dilations (or more general group, which acts on the space of solutions) 
is inside the approach
(in wavelet case), this
method resembles the action of a microscope. We have contribution to
final result from each scale of resolution from the whole underlying 
infinite scale of spaces. 
In part 4 we consider numerical modelling  of
Wigner functions,  
which explicitly demonstrates quantum interference of
generalized coherent states.

\section{WIGNER EQUATIONS}
According to Weyl transform, quantum state (wave function or density operator) corresponds
to Wigner function, which is the analog of classical phase-space distribution [2].
We consider the following form of differential equations for time-dependent WF, $W=W(p,q,t)$:
\begin{eqnarray}
W_t=\frac{2}{\hbar}\sin\Big[\frac{\hbar}{2}
(\partial^H_q\partial^W_p-\partial^H_p\partial^W_q)\Big]\cdot HW
\end{eqnarray}
Let 
$$\hat{\rho}=|\Psi_\epsilon><\Psi_\epsilon|$$
be the density operator or projection operator corresponding to the energy eigenstate
$|\Psi_\epsilon>$ with energy eigenvalue $\epsilon$. Then time-independent Schroedinger equation
corresponding to Hamiltonian
\begin{eqnarray}
\hat{H}(\hat{p},\hat{q})=\frac{\hat{p}^2}{2m}+U(\hat{q}),
\end{eqnarray}
where $U(\hat{q})$ is an arbitrary polynomial function (related beam dynamics models
considered in [3]-[14]) on $\hat{q}$, is [2]:
\begin{eqnarray}
\hat{H}\hat{\rho}=\epsilon\hat{\rho}
\end{eqnarray}
After Weyl-Wigner mapping we get the following equation on WF in c-numbers:
\begin{eqnarray}
H\big(p+\frac{\hbar}{2i}\frac{\partial}{\partial q},q-\frac{\hbar}{2i}\frac{\partial}{\partial p}\Big)W(p,q)
  =\epsilon W(p,q)
\end{eqnarray}
or
\begin{eqnarray*}
&&\Big( \frac{p^2}{2m}+\frac{\hbar}{2i}\frac{p}{m}\frac{\partial}{\partial q}-
 \frac{\hbar^2}{8m}\frac{\partial^2}{\partial q^2}\Big)W(p,q)+\\
&& U\Big(q-\frac{\hbar}{2i}\frac{\partial}{\partial p}\Big)W(p,q)=\epsilon W(p,q)
\end{eqnarray*}
 After expanding the potential $U$ into the Taylor series we have two real partial differential equations. 
In the next section we consider  variation-wavelet approach for the solution of these 
equations for the case of an arbitrary polynomial $U(q)$, which corresponds to a finite number 
of terms in equations (7) up to any finite order of $\hbar$.

\section{VARIATIONAL MULTISCALE REPRESENTATION}

Let L be an arbitrary (non)linear differential operator with matrix dimension $d$, 
which acts on some set of functions
$\Psi\equiv\Psi(x,y)=\Big(\Psi^1(x,y),\dots,\Psi^d(x,y)\Big), \quad x,y\in\Omega\subset R^2$
from $L^2(\Omega)$:
\begin{equation}
L\Psi\equiv L(Q,x,y)\Psi(x,y)=0,
\end{equation}
where
$Q\equiv Q(x,y,\partial /\partial x,\partial /\partial y)$.

Let us consider now the N mode approximation for solution as the following ansatz (in the same way
we may consider different ansatzes):
\begin{equation}
\Psi^N(x,y)=\sum^N_{r,s=1}a_{rs}\Psi_r(x)\Phi_s(y)
\end{equation}
We shall determine the coefficients of expansion from the following variational conditions
(different related variational approaches are considered in [3]-[14]):
\begin{equation}
\ell^N_{k\ell}\equiv\int(L\Psi^N)\Psi_k(x)\Phi_\ell(y)\ud x\ud y=0
\end{equation}
So, we have exactly $dN^2$ algebraical equations for  $dN^2$ unknowns $a_{rs}$.
But in the case of equations for WF (7) we have overdetermined system of equations:
$2N^2$ equations for $N^2$ unknowns $a_{rs}$ (in this case $d=1$).
In this paper we consider non-standard method for resolving this problem,
which is based on biorthogonal wavelet expansion. So, instead of expansion (9) we consider
the following one:
$$
\Psi^N(x,y)=
\sum^N_{r,s=1}a_{r,s}\Psi_r(x)\Psi_s(y)+\sum^N_{i,j=1}\widetilde{a}_{ij}
\widetilde{\Psi}_i(x)\widetilde{\Phi}_j(y)
$$
where $\widetilde{\Psi}_i(x), \widetilde{\Phi}_j(y)$ are the bases dual to initial ones.

Because wavelet functions are the generalization of coherent states we consider 
an expansion on this overcomplete set of basis wavelet functions as a generalization of 
standard coherent states expansion.
So, variational approach reduced the initial problem (8) to the problem of solution 
of functional equations at the first stage and some algebraical problems at the second
stage. 

We  consider the multiresolution expansion as the second main part of our 
construction. 
We have contribution to        
final result from each scale of resolution from the whole
infinite scale of increasing closed subspaces $V_j$:
$$
...V_{-2}\subset V_{-1}\subset V_0\subset V_{1}\subset V_{2}\subset ...
$$

The solution is parametrized by solutions of two reduced algebraical
problems, one is linear or nonlinear (10)
(depends on the structure of operator L)  and the others are some linear
problems related to computation of coefficients of algebraic equations (10).
These coefficients can be found  by some wavelet methods.
We use compactly supported wavelet basis functions for expansions (9).
We may consider different types of wavelets including general wavelet packets.

Now we concentrate on the last additional problem, that comes from 
overdeterminity of equations (7), which demands to consider biorthogonal wavelet expansions.
It leads to equal number of equations and 
unknowns in reduced algebraical system of equations (10).
We start with two hierarchical sequences of approximations spaces:
$$
\dots V_{-2}\subset V_{-1}\subset V_{0}\subset V_{1}\subset V_{2}\dots,$$
$$\dots \widetilde{V}_{-2}\subset\widetilde{V}_{-1}\subset
\widetilde{V}_{0}\subset\widetilde{V}_{1}\subset\widetilde{V}_{2}\dots, 
$$
and as usually,
$W_0$ is complement to $V_0$ in $V_1$, but now not necessarily orthogonal
complement.
Functions $\varphi, \tilde\varphi$ generate a multiresolution analysis.
$\varphi(x-k)$, $\psi(x-k)$ are synthesis functions,
$\tilde\varphi(x-\ell)$, $\tilde\psi(x-\ell)$ are analysis functions.
Synthesis functions are biorthogonal to analysis functions. Scaling spaces
are orthogonal to dual wavelet spaces.
Biorthogonal point of view is more flexible and stable under the action of large class
of operators while orthogonal (one scale for multiresolution) is fragile,
all computations are much more simple and we accelerate the rate of convergence of our 
expansions (9). By analogous anzatzes and approaches we may construct also the 
multiscale/multiresolution representations for solution of time dependent Wigner equation (4) [14].

\begin{figure}[htb]
\centering
\includegraphics*[width=65mm]{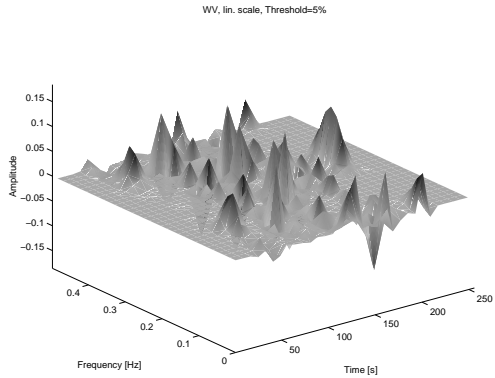} 
\caption{Wigner function for 3 wavelet packets.}
\end{figure}
\begin{figure}[htb]
\centering
\includegraphics*[width=65mm]{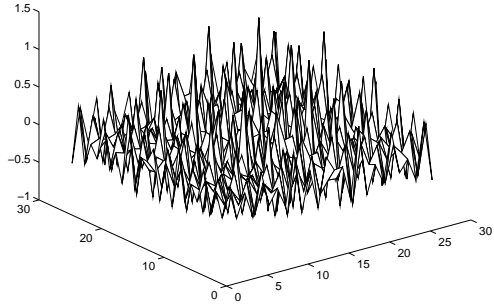}
\caption{Multiresolution/multiscale representation for Wigner function.}
\end{figure}

\section{NUMERICAL MODELLING}

So, our constructions give us the following N-mode representation for solution of Wigner equations
(7):
\begin{equation}
W^N(p,q)=\sum^N_{r,s=1}a_{rs}\Psi_r(p)\Phi_s(q)
\end{equation}
where $\Psi_r(p)$, $\Phi_s(q)$ may be represented by some family of (nonlinear)
eigenmodes with the corresponding multiresolution/multiscale representation in the
high-localized wavelet bases: 

\begin{eqnarray*}
\Psi_k(p)&=&\Psi^{M_1}_{k, slow}(p)+\sum_{i\ge M_1}\Psi^i_k(\omega_i^1p),
\quad \omega^1_i \sim 2^i\\
\Phi_k(q)&=&\Phi^{M_2}_{k, slow}(q)+\sum_{j\ge M_2}\Phi^j_k(\omega_j^2q),
\quad \omega^2_j \sim 2^j
\end{eqnarray*}
Our (nonlinear) eigenmodes are more realistic for the modelling of 
nonlinear classical/quantum dynamical process  than the corresponding linear gaussian-like
coherent states. Here we mention only the best convergence properties of expansions 
based on wavelet packets, which  realize the so called minimal Shannon entropy property.
On Fig.~1 we present the numerical modelling [15] of Wigner function for a simple model of beam motion,
which explicitly demonstrates quantum interference property. On Fig.~2 we present 
the multiscale/multiresolution representation (11) for solution of Wigner equation.

\section{ACKNOWLEDGMENTS}

We would like to thank The U.S. Civilian Research \& Development Foundation (CRDF) for
support (Grants TGP-454, 455), which gave us the possibility to present our nine papers during
PAC2001 Conference in Chicago and Ms.Camille de Walder from CRDF for her help and encouragement.

 \end{document}